\documentclass[preprint,12pt]{elsarticle}

\usepackage{amssymb}
\usepackage{amsmath}

\usepackage{dsfont} 
\usepackage{hyperref}
\usepackage{tabularray}
\usepackage{multirow}
\usepackage{colortbl}
\usepackage{booktabs}
\usepackage{graphicx}   
\usepackage{subcaption} 
\usepackage{cleveref} 
\journal{Computers in Biology and Medicine} 

\makeatletter
\def\ps@pprintTitle{%
 \let\@oddhead\@empty
 \let\@evenhead\@empty
 \let\@oddfoot\@empty
 \let\@evenfoot\@empty
}
\makeatother

\begin{document}

\begin{frontmatter}



\title{Validation of Conformal Prediction in Cervical Atypia Classification} 

\author[dst_liu,cmiv_liu]{Misgina Tsighe Hagos}
\author[fimm_uoh]{Antti Suutala}
\author[fimm_uoh]{Dmitrii Bychkov}
\author[fimm_uoh]{Hakan Kücükel}
\author[fimm_uoh,dwch_uu,dgph_ki]{Joar von Bahr}
\author[dst_liu,cmiv_liu]{Milda Poceviciute}
\author[fimm_uoh,dgph_ki]{Johan Lundin}
\author[fimm_uoh,dwch_uu]{Nina Linder}
\author[dst_liu,cmiv_liu,sectra]{Claes Lundström}

\affiliation[dst_liu]{organization={Department of Science and Technology},
            addressline={Linköping University},
            city={Norrköping},  
            country={Sweden}}

\affiliation[cmiv_liu]{organization={Center for Medical Imaging Science and Visualization},
            addressline={Linköping University},
            city={Linköping},
            country={Sweden}}
\affiliation[fimm_uoh]{organization={Institute for Molecular Medicine Finland - FIMM},
            addressline={University of Helsinki},
            city={Helsinki},
            country={Finland}}
\affiliation[dwch_uu]{organization={Department of Women's and Children's Health},
            addressline={Uppsala University},
            city={Uppsala},
            country={Sweden}}
\affiliation[dgph_ki]{organization={Department of Global Public Health},
            addressline={Karolinska Institutet},
            city={Stockholm},
            country={Sweden}}
\affiliation[sectra]{organization={Sectra AB},
            city={Linköping},
            country={Sweden}}



\begin{abstract}

Deep learning based cervical cancer classification can potentially increase access to screening in low-resource regions. However, deep learning models are often overconfident and do not reliably reflect diagnostic uncertainty. Moreover, they are typically optimized to generate maximum-likelihood predictions, which fail to convey uncertainty or ambiguity in their results. Such challenges can be addressed using conformal prediction, a model-agnostic framework for generating prediction sets that contain likely classes for trained deep-learning models. The size of these prediction sets indicates model uncertainty, contracting as model confidence increases. However, existing conformal prediction evaluation primarily focuses on whether the prediction set includes or covers the true class, often overlooking the presence of extraneous classes. We argue that prediction sets should be truthful and valuable to end users, ensuring that the listed likely classes align with human expectations rather than being overly relaxed and including false positives or unlikely classes. In this study, we comprehensively validate conformal prediction sets using expert annotation sets collected from multiple annotators. We evaluate three conformal prediction approaches applied to three deep-learning models trained for cervical atypia classification. Our expert annotation-based analysis reveals that conventional coverage-based evaluations overestimate performance and that current conformal prediction methods often produce prediction sets that are not well aligned with human labels. Additionally, we explore the capabilities of the conformal prediction methods in identifying ambiguous and out-of-distribution data.

\end{abstract}


\begin{highlights}
\item We perform the first validation of conformal prediction methods in cervical atypia classification using annotation sets collected from multiple experts.
\item Conventional evaluation metrics of conformal prediction sets generate overestimated performances when compared against expert annotation set-based validations.
\item Conformal prediction sets are better suited for capturing aleatoric uncertainty (caused by data ambiguity) rather than epistemic uncertainty (caused by Out-of-Distribution data)

\end{highlights}


\end{frontmatter}



\section{Introduction}
\label{section:introduction}

Cervical cancer is the fourth most prevalent cancer among women worldwide and has caused approximately 350,000 deaths in 2022, with over 90\% of these occurring in low- and middle-income countries \cite{sung2021global,WHO2024}. A key contributor to the high mortality rate is the lack of access to cervical cancer screening, which limits early detection and timely intervention \cite{sung2021global}. Analyzing with microscopy of Papanicolaou-stained cytology samples, i.e. Pap smears, is one of the recommended methods for cervical cancer screening \cite{IARC2022}. Advancements in digital pathology and deep learning enable Artificial Intelligence (AI) supported analysis of Papanicolaou smears with high accuracy and have the potential to increase access to screening in regions with a shortage of pathologists \cite{holmstrom2021point}. The Bethesda system is recommended for reporting the results of Pap smear analysis and contains categories for squamous cell atypia: negative for intraepithelial lesion or malignancy (NILM), low-grade squamous
intraepithelial lesion (LSIL), and high-grade squamous intraepithelial lesion (HSIL), which align with the typical way deep learning models output predictions, including the most probable category and an associated probability \cite{alrajjal2021squamous}. The Bethesda system also accounts for ambiguous findings by categorising atypical squamous cells of unknown significance (ASC-US) and atypical squamous cells that can not exclude HSIL (ASC-H). However, deep learning models often produce overconfident and poorly calibrated probability outputs, which do not reliably reflect diagnostic uncertainty \cite{guo2017calibration,nixon2019measuring}. Addressing this limitation is crucial for building transparent and informative models in cervical cancer screening.




Conformal prediction is one of many approaches to formulating uncertainty in deep learning models. It generates a prediction set that covers the true class with a high probability \cite{vovk2005algorithmic}. For input data $x \in X$, conformal prediction generates a prediction set of the most likely $k$ classes $\{y_1, ... , y_k\} \subseteq Y$, where $Y$ is the set of all possible classes. This way, it enables end users to know which category to rule out and which to rule in. The size of the prediction set is dynamic and depends on the model's confidence in its outputs. So, a larger prediction set size implies uncertain outputs,  while a smaller set indicates higher certainty \cite{angelopoulos2021gentle}.

Conformal prediction has been widely applied across domains, from generic datasets to high-stakes medical classification tasks \cite{clark2024conformal}. However, its evaluation primarily depends on metrics such as classification coverage, size-stratified coverage, and set width \cite{angelopoulosuncertainty}. These metrics assess whether the generated conformal prediction sets satisfy their fundamental requirement—ensuring that the true class is included within the prediction set, i.e.:

\begin{equation*}
    y \in \{y_1, ... , y_k\} \ \forall (x,y) \in X \times Y
\end{equation*}
    
\begin{figure}[h]
    \centering
    \includegraphics[width=1\textwidth]{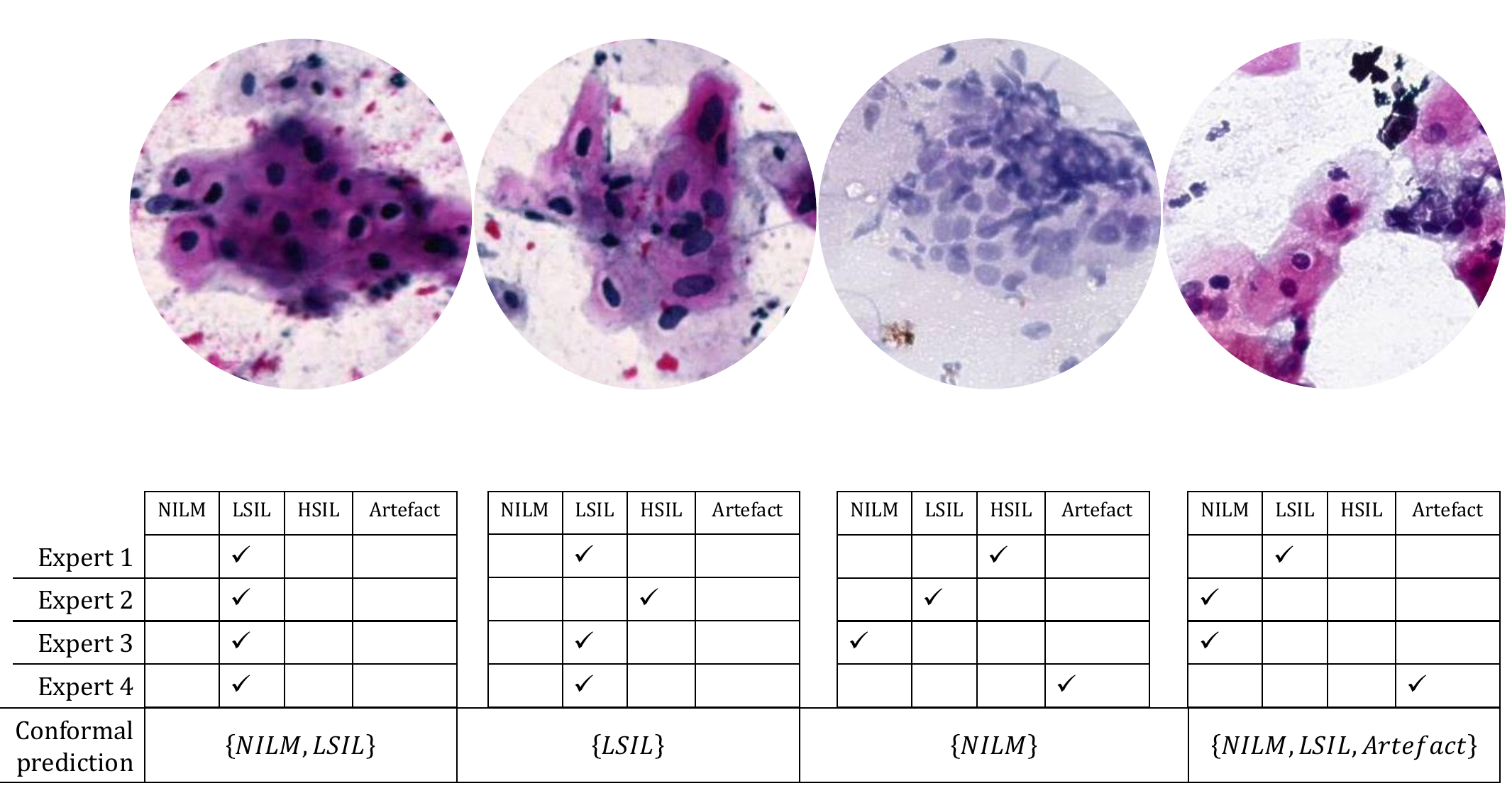}

    \caption{Sample tiles, with their corresponding annotations from four expert annotators and conformal prediction sets generated for a trained model. The output prediction sets perfectly cover one or more of the experts' annotations. However, they also usually add extraneous labels and fail to mirror the disagreement between annotators, as seen in the first three examples. The last prediction set correctly outputs the experts' annotations.} 
    \label{figure:sample_tiles_annotations_and_cp}
\end{figure}

In real-world applications, prediction sets are presented to end users as the most likely model outputs. Thus, it is essential to ensure that these sets are both concise and informative while accurately representing classes with high likelihood. A major limitation of current approaches is that prediction sets can contain uninformative or extraneous labels, potentially misleading users. In addition, prediction sets might fail to correctly include all the likely classes that are annotated by experts. \Cref{figure:sample_tiles_annotations_and_cp} illustrates this issue, where the prediction sets of the first three example inputs only cover one or more of the expert-annotated true classes and sometimes add irrelevant labels. On the contrary, the last example in \Cref{figure:sample_tiles_annotations_and_cp} presents an ideal conformal prediction set output that contains all expert-annotated classes. This highlights a gap in the validation of conformal prediction, as their overall coverage (i.e., the true class appearing in a prediction set) does not guarantee that individual elements within these sets are themselves meaningful.


Comprehensive validation of prediction sets remains challenging due to the lack of fully annotated datasets with multiple expert labels for each test sample. To address this gap, we collected multi-annotator labels for a cervical cytology dataset and systematically evaluated three conformal prediction approaches across three deep-learning models. Additionally, we assessed each method’s ability to capture the two primary sources of uncertainty: aleatoric uncertainty, which arises from inherent noise or ambiguity in the training data, and epistemic uncertainty, which reflects the model's lack of knowledge \cite{hora1996aleatory,hullermeier2021aleatoric}. The ability to capture aleatoric uncertainty was evaluated based on how well each method accounted for existing ambiguity within the dataset, while we gauged the capacity to capture epistemic uncertainty by testing the ability to identify out-of-distribution (OOD) samples.


In this paper, we present the following key contributions,

\begin{itemize}
    \item Using our custom-built platform, we collect expert annotations of tiles.
    \item We train three deep-learning models for classifying cervical atypia into one of four categories and generate their prediction sets using three conformal prediction approaches.
    \item We perform the first validation of conformal prediction sets using annotation sets collected from four experts. Our validation shows that conventional evaluation metrics of conformal prediction appear to yield overestimated performance assessments and that conformal prediction sets do not align well with the annotation sets of the experts.
    \item We evaluate the performance of the prediction sets in capturing aleatoric (ambiguous data) and epistemic uncertainty (OOD data) and show that the prediction sets perform better at capturing ambiguity in the dataset than detecting OOD data.
\end{itemize}


\section{Methods}
\label{section:methods}

In this section, we provide details of our data collection process, model training strategies, conformal prediction methods, and performance assessment approaches.

\subsection{Data collection}
\label{section:methods:data_collection}

In this study, we used 301 conventional Pap smears collected from 294 women at the Kinondo Kwetu Hospital in rural Kenya (Kinondo, Kwale County). The smears were digitized using a portable whole-slide microscope scanner (Grundium Ocus) equipped with a 20$\times$ objective and a numerical aperture of 0.40, producing digitized images with a pixel size of 0.48 $\mu m/pixel$. The digitized slides measured approximately 100,000$\times$50,000 pixels, corresponding to the dimensions of the sample area of a standard microscope glass slide (25 mm$\times$ 50 mm). Sample preparation and processing are described in detail in a previous publication \cite{holmstrom2021point}.

\begin{figure}
    \centering
    \includegraphics[width=1\linewidth]{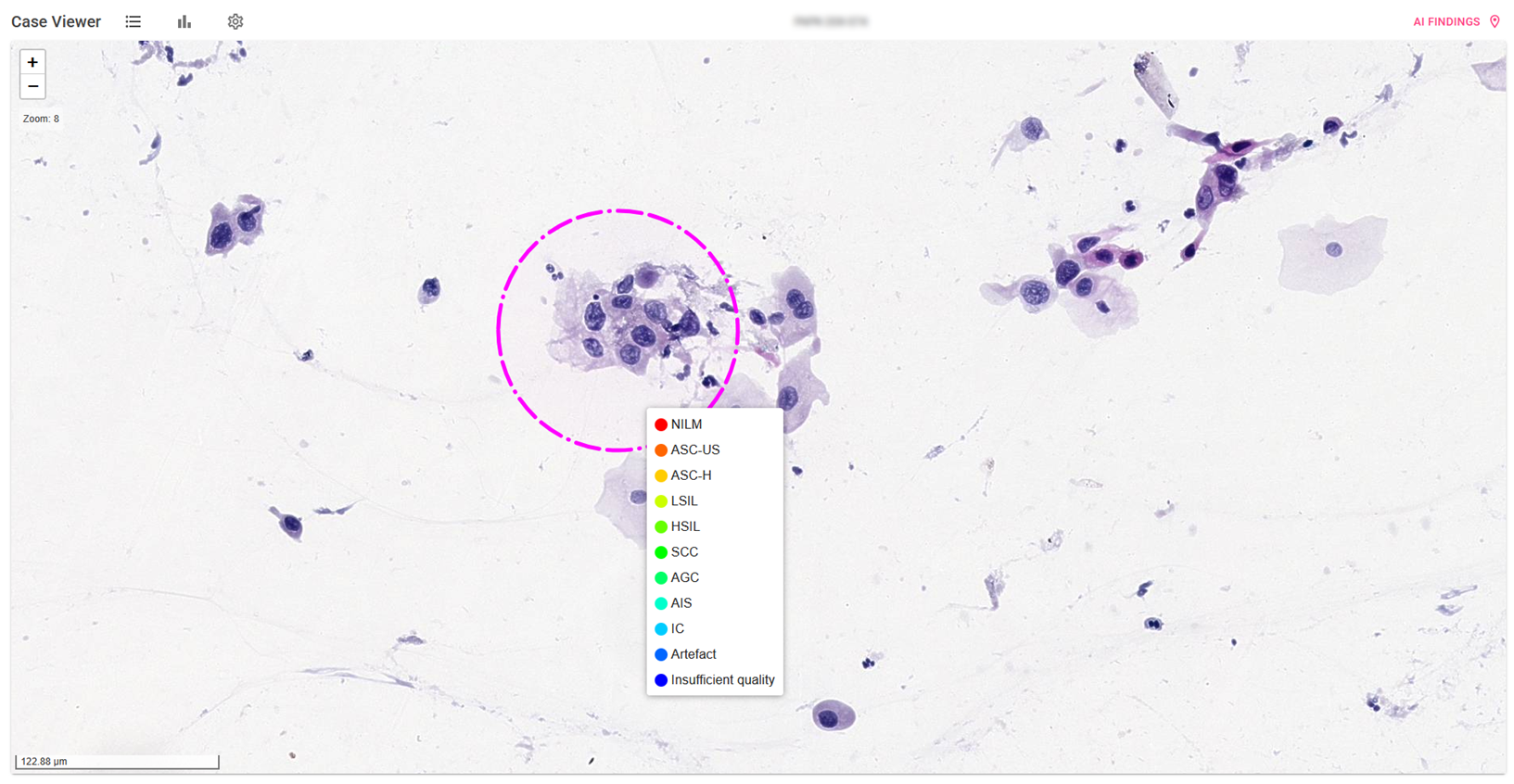}
    \caption{User interface of the web-based annotation platform for labelling AI-generated regions of interest (ROIs) on whole slide images. A specific ROI is highlighted, and a predefined list of options offers the diagnostic categories, such as NILM, LSIL, HSIL, and Artefact. 
    }
    \label{figure:web_based_annotation_platform_screengrab}
\end{figure}

To collect annotations from experts, we developed a platform that provides secure remote access to the data and a web-based application with a user interface designed to help experts browse whole slide images (WSIs), visualize AI-generated regions of interest (ROI), and label them. Experts can review AI-generated ROI, zoom in and out on the WSI to view the surrounding area and select the label that best describes it. For Pap smears, the options are NILM, ASC-US, ASC-H, LSIL, HSIL, Squamous Cell Carcinoma (SCC), Atypical Glandular Cells (AGC), Adenocarcinoma in Situ (AIS), Invasive Carcinoma (IC), Artefact, and Insufficient quality. Once annotations are made, they are securely submitted and stored in real-time. The application is developed with open-source components and incorporates secure data management through role-based access control. It utilizes containerization technologies (Docker, Docker Inc, Palo Alto, CA) and cloud services (Azure DevOps, Microsoft Corp, Redmond, WA) to support continuous improvements based on expert feedback. Additionally, serverless functions and web APIs facilitate secure data import and export, ensuring reliability through Microsoft Azure resources. A screengrab of the application's user interface is shown in \Cref{figure:web_based_annotation_platform_screengrab}.

We select a subset of tiles for a given WSI to avoid overwhelming the expert annotators. An AI model was developed and trained using a cloud-based deep learning platform \footnote{\url{https://www.aiforia.com/}} (Aiforia Create, Aiforia Technologies, Helsinki, Finland) to generate an initial set of ROI. The process follows the principles outlined in the model training and development details, which have been published previously \cite{holmstrom2021point}. The AI model detects areas of low-grade squamous intraepithelial lesions (LSILs) and high-grade squamous intraepithelial lesions (HSILs) in Pap smear WSIs and segments them into an ROI. In this study, these regions were converted into tiles by calculating the centroid of the polygon contouring the area and resizing the tile to a fixed size of 144 × 144 $\mu m^2$ (300 $\times$ 300 pixels). Overlapping boxes were removed by selecting the one with the highest score. A maximum of 24 tiles were selected from each slide based on the highest confidence scores and subsequently presented on the annotation platform. High-confidence tiles from each category were prioritized to identify and include relevant diagnostic characteristics. This number of tiles kept the review workload manageable, enabling thorough evaluation without overwhelming the experts.


Using the AI model, a total of 7,119 individual tiles were identified and presented from 301 WSIs, with each WSI contributing between 5 and a maximum of 24 tiles on the annotation platform. The mean number of tiles per WSI was 23.7. The AI model aimed to identify up to 24 suitable tiles per WSI; however, when fewer than 24 were identified, all available tiles from that WSI were used. Six cytology experts participated in the annotation process. Participation levels ranged from 8.2\% to 100\%, with a mean of 49.8\%.


To reduce the variability of the ground truth of the collected annotation, we used majority voting and only included tiles where at least two experts agreed on the label/class (n = 5,913). Furthermore, within the scope of this study, we narrowed the inclusion criteria to cover tiles labelled as NILM, LSIL, HSIL, or Artefact classes. Here, we define ‘LSIL’ as a combination of the LSIL and ASC-US categories and ‘HSIL’ as a combination of the ASC-H, HSIL, and SCC categories. This data curation process formed the basis of the dataset, which contained 5,239 individual tiles from 299 WSIs.

We divided the dataset into training, calibration, and test datasets. The training and calibration set contained 4,801 individual tiles from 274 WSIs and was split using a 70/30 ratio. The calibration set is intended for use in conformal prediction (See details in \Cref{section:conformal_prediction_approachs}). The test set was formed from a separate group of 25 WSIs, where a designated group of four experts had fully completed the annotation assignments. The test set comprised a total of 438 tiles. We ensured that there was no WSI overlap between the sets. Details of the datasets are shown in \Cref{table:data_distribution}.

\begin{table}
\centering
\caption{The distribution of data across the training, calibration, and test sets, detailing the number of tiles allocated to each subset, grouped by label.}
\begin{tblr}{
  hline{1,6} = {-}{0.08em},
  hline{2} = {-}{},
}
Label    & Training & Calibration & Test \\
NILM     & 2151     & 923         & 264  \\
LSIL     & 740      & 318         & 77   \\
HSIL     & 123      & 52          & 39   \\
Artefact & 346      & 148         & 58   
\end{tblr}
\label{table:data_distribution}
\end{table}

For a single tile, $x \in X$, where $X$ is the set of all test set tiles, there are two versions of its ground truth label, $y$. These are:

\begin{enumerate}
    \item Annotation sets, which are annotations collected from the four experts, where each expert provides one label for each tile. We represent this ground truth as $y^0$.
    \item Per-tile consensus of the expert annotations, which is computed using majority voting. This contains only one label per tile and is represented as $y^1$.
\end{enumerate}

\subsection{Model training}
\label{section:methods:model_training}

We trained the classifier heads of the deep learning models ResNet-18 \cite{he2016deep}, ResNet-50 \cite{he2016deep}, and EfficientNet-B0 \cite{tan2019efficientnet} models to classify extracted tiles into the four categories. The models were pre-trained on ImageNet \cite{russakovsky2015imagenet}. We used a batch size of 32, with the training process set to run for a maximum of 85 epochs. We used an Adam optimizer \cite{kingma2015adam} with a learning rate initialized at $1e-3$ and dynamically adjusted using the ReduceLROnPlateau learning rate scheduler. To improve computational efficiency, mixed precision (16-bit) training was utilized.


The training set was augmented with multiple techniques, including random horizontal flipping, colour jitter (with brightness, contrast, saturation, and hue adjustments of $\pm$0.2, $\pm$0.2, $\pm$0.2, and $\pm$0.1, respectively), and random affine transformations (rotation of $\pm$10 degrees, translation of $\pm$10\%, and scaling of $\pm$10\%). All images were resized to 224$\times$224 pixels and normalized using the ImageNet mean [0.485, 0.456, 0.406] and standard deviation [0.229, 0.224, 0.225] values.

On the test set, the ResNet50, ResNet18, and EfficientNet-B0 models achieved an AUC of 0.88, 0.90, 0.87, respectively. We use these trained models to generate conformal prediction sets of the test set tiles.

\subsection{Conformal prediction approaches}
\label{section:conformal_prediction_approachs}

For our input tiles, $x \in X$, a trained model outputs softmax outputs for each class, $\hat{f}(x)\in [0, 1]^K$, where $K=4$ is the total number of categories in our dataset. In conformal prediction, we are interested in constructing a prediction set $C$, for test set tiles, containing all possible classes $C(X_{test}) \subset {1, ... , K}$, that satisfies,

\begin{equation}
    C(X_{\text{test}}) = \{ y : \hat{f}(X_{\text{test}})_y \geq 1 - \hat{q} \},
\end{equation}

\noindent where $\hat{q}$ is a threshold value computed using a calibration set $(X_{cal}, Y_{cal})$. We use three conformal prediction approaches to generate prediction sets for all the models: the least ambiguous set-valued classifier (LAC), adaptive prediction sets (APS), and regularized adaptive prediction sets (RAPS). They are explained in detail next.

\subsubsection{Least ambiguous set-valued classifier}

LAC \cite{sadinle2019least} first computes a score function, $s$,

\begin{equation}
    s(x,y) = 1 - \hat{f}(x)_{y_i},
\end{equation}

\noindent where $y_i$ is the index of the true class. This would output ${s_1, ..., s_n}$ for each element of the calibration set. For a user defined error rate, $\alpha$, we then compute the quantile,

\begin{equation}
    \hat{q} = quantile \left( \{s_1, \dots, s_n\}; \frac{ \left\lceil (1 - \alpha)(n + 1)\right\rceil}{n}  \right)
\end{equation}

\noindent where $(\frac{n+1}{n})$ is a finite sample size correction. We can then construct the prediction set as follows,

\begin{equation}
    C(x_{\text{test}}) = \{ y : s(x_{\text{test}}, y_{\text{test}}) \leq \hat{q} \} = \{ y : \hat{f}(x_{\text{test}})_y \geq 1 - \hat{q} \}
\end{equation}

Since it only takes the softmax values of the true class when computing the quantile, LAC can result in empty prediction sets for uncertain cases.

\subsubsection{Adaptive prediction sets}

APS \cite{romano2020classification}first sorts the softmax outputs, $\hat{f}(x)$, in descending order, giving us a list of softmax values $\pi(x)$. The score function is computed by taking the sum of the softmax values in $\pi(x)$ from the start index to the true class's index.

\begin{equation}
    s(x,y) = \sum_{j=1}^{k} \hat{f}(x)_ {\pi_{j}(x)}, 
\end{equation}

\noindent where $\pi_k(x)$ represents the true class. The quantile $\hat{q}$ is computed similar to LAC. Now that we have a score function $s$ for all the test tiles, we can generate the conformal prediction sets as follows,

\begin{equation}
    C(x_{test}) = \{\pi_1, ..., \pi_k\}, k = \inf \left\{ k : \sum_{j=1}^{k} \hat{f}(x_{test}) \pi_{j}(x) \geq \hat{q} \right\}
\end{equation}

\noindent Even if APS encounters uncertain cases since it takes the summation of softmax values until they exceed $\hat{q}$, it successfully overcomes LAC's issue of generating empty sets.

\subsubsection{Regularized adaptive prediction sets}

Even though APS generates non-empty sets, it can lead to relatively larger sets. RAPS builds on APS by incorporating regularization, ensuring coverage while adjusting prediction set sizes based on model uncertainty \cite{angelopoulosuncertainty}. Conformity scores in RAPS are computed as follows,

\begin{equation}
    s(x,y) = \sum_{j=1}^{k} \hat{f}(x)_ {\pi_{j}} + \lambda(k - k_{reg}),
\end{equation}

\noindent where $\pi_k$ represents the true class and $k_{reg}$ is the optimal prediction set size. This is achieved using a subset of the calibration set to optimize for the hyperparameter $\lambda$. Following \citet{angelopoulosuncertainty}, we search for  $\lambda \in \{0.001, 0.01, 0.1, 0.2, 0.5 \}$ using 20\% of the calibration set.

The quantile $\hat{q}$ is computed as in LAC and the prediction set is determined as follows,

\begin{equation}
    C(x_{test}) = \{\pi_1, ..., \pi_k\}, k = \inf \left\{ k : \sum_{j=1}^{k} \hat{f}(x_{test}) \pi_{j} + \lambda (k - k_{reg}) \geq \hat{q} \right\}
\end{equation}



\subsection{Evaluation}
\label{section:evaluation_metrics}

We employ two distinct methodologies to assess the generated prediction sets. The first approach assesses their alignment with human experts by utilizing the experts' annotation sets and their consensus as ground truth. The second evaluates the prediction sets' effectiveness in detecting data ambiguity and out-of-distribution (OOD) data.

\subsubsection{Alignment with expert annotations}

Here, we use evaluation metrics to directly assess the performance of the generated prediction sets using ground truth labels. This is done in two ways based on the two versions of our ground truth labels (expert annotation sets and their consensus) as follows,


\begin{enumerate}
    \item Coverage-based validation, where we assess if prediction sets cover input tiles' consensus ground truth label within a reasonable set width. This includes the conventional metrics: classification coverage (CC), size-stratified coverage (SSC), and mean width. For an input tile $x_i$, a prediction set output $\mathcal{C}(x_i)$ and its corresponding consensus groundtruth label $y^1_i$, the coverage-based metrics are defined as follows,

    \begin{equation} 
        \text{CC} = \frac{1}{N} \sum_{i=1}^{N} \mathds{1} \left\{ y^1_i \in \mathcal{C}(x_i) \right\}
    \end{equation}

    \begin{equation} 
        \text{SSC} = \min_{g \in \{1, \dots, G\}} \frac{1}{|I_g|} \sum_{i \in I_g} \mathds{1} \left\{ y^1_i \in \mathcal{C}(x_i) \right\}
    \end{equation}

\noindent where tiles with similar prediction set sizes are grouped into $g$ bins, $I_g$ is a group of sets that belong to the $g^{th}$ size group, and $G$ is the number of distinct size groups.

    \begin{equation}
        \text{Mean\ width} = \frac{1}{N} \sum_{i=1}^{N} |\mathcal{C}(x_i)|
    \end{equation}
    
    \item Agreement-based validation, which assesses the quality of individual conformal prediction sets in mirroring experts' annotation sets. The metrics mean precision, mean recall, mean F1 score, and mean Jaccard coefficient are used. Even though these are well-established metrics, they have not been used in the literature to evaluate conformal prediction sets. We believe this is due to the challenges in collecting an expert annotation set where multiple human experts' labels are provided per a single test input. Given a conformal prediction set output $\mathcal{C}(x_i)$ and the corresponding expert annotation set ground truth $y^0_i$ for a tile $x_i$, we define the agreement-based validation metrics as follows:


    \begin{equation}
        \text{Mean\ Precision} = \frac{1}{N} \sum_{i=1}^{N} \frac{|\mathcal{C}(x_i) \cap y^0_i|}{|\mathcal{C}(x_i)|}
    \end{equation}

    
    \begin{equation}
        \text{Mean\ Recall} = \frac{1}{N} \sum_{i=1}^{N} \frac{|\mathcal{C}(x_i) \cap y^0_i|}{|y^0_i|}
    \end{equation}
    
    
    \begin{equation}
        \text{Mean\ F1\ score} = \frac{1}{N} \sum_{i=1}^{N} \left(2 \times \frac{\text{Precision}_i \times \text{Recall}_i}{\text{Precision}_i + \text{Recall}_i}\right )
    \end{equation}
    
    
    \begin{equation}
        \text{Mean\ Jaccard\ Coefficient} = \frac{1}{N} \sum_{i=1}^{N} \frac{|\mathcal{C}(x_i) \cap y^0_i|}{|\mathcal{C}(x_i) \cup y^0_i|}
    \end{equation}

\end{enumerate}

For the alignment-with-experts-based evaluation, we used four alpha values $\alpha \in \{0.05, 0.1, 0.15$, $0.2\}$ to generate the conformal prediction sets. We then selected the best-performing alpha value for the rest of the evaluations.

\subsubsection{Capturing aleatoric and epistemic uncertainty}

In addition to the evaluation metrics, we assess the performance of the generated prediction sets in identifying aleatoric and epistemic uncertainty. To evaluate prediction sets' ability to capture aleatoric uncertainty, we assess whether their uncertainty correctly mirrors the experts' uncertainty in annotating the tiles. For each tile, we compare whether the count of unique labels in the experts' annotation set is the same as the set width of the prediction set generated for said tile. Note that high model uncertainty is reflected in increased prediction set width.

To assess conformal prediction's capability in capturing epistemic uncertainty, we use two OOD data types: the first OOD data is created by noising our test set and the second OOD data is a bone marrow cytomorphology dataset \cite{matek2021bone} that our models did not previously see. The second OOD data, the bone marrow dataset, contains 171381 tiles from bone marrow smears of 945 patients, stained using the May-Grünwald-Giemsa/Pappenheim stain. The images were acquired with a brightfield microscope at 40× magnification under oil immersion.

\begin{figure}[h]
    \centering
    \includegraphics[width=1.0\textwidth]{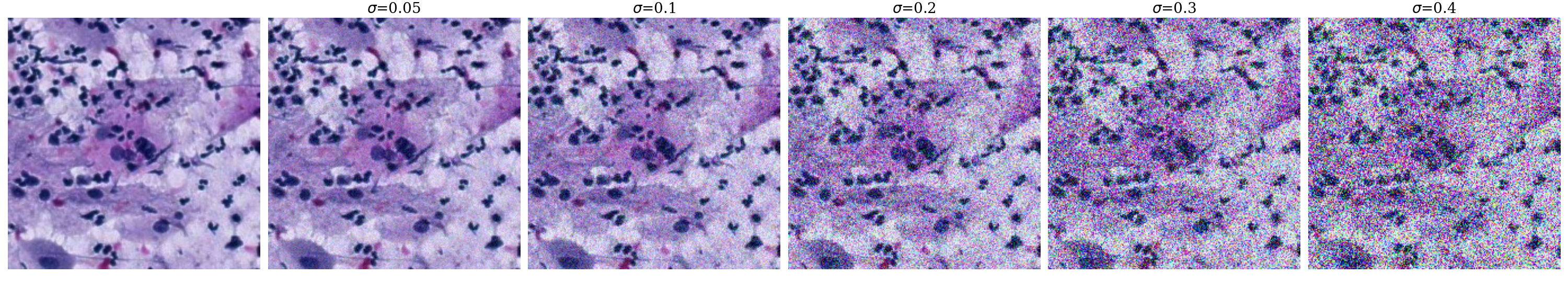}
    \caption{A sample input tile (shown on the left) and its Out-of-Distribution variants generated by adding a Gaussian noise at different $\sigma$ values.}
    \label{figure:sample_ind_to_ood_tile}
\end{figure}

We generate the first OOD data by introducing Gaussian noise into the test dataset. Here, we are interested in whether increasing noise levels would increase uncertainty in the generated conformal prediction sets. Given an input test tile $I$, we obtain a noised image $I'$ as follows:

\[
I' = I + \mathcal{N}(0, \sigma)
\]

\noindent where $\mathcal{N}(0, \sigma)$ represents Gaussian noise with zero mean and standard deviation  $\sigma$. To systematically assess OOD behavior, we generate multiple OOD variants of the in-distribution (InD) tile images using different $\sigma$ values. A sample input tile and its OOD variants are plotted in \Cref{figure:sample_ind_to_ood_tile}.

\section{Results}
\label{section:results}

Inter-rater reliability analysis using Fleiss' Kappa yielded $k$ = 0.32 (95\% CI: 0.27-0.36), indicating fair agreement \citep{fleiss1971measuring,landis1977measurement}. This also suggests variability in annotator decisions, potentially due to tile ambiguity.

\subsection{Performance in predicting experts' annotations}
\label{section:results:predicting_experts_annotation}

\begin{figure}[!t]
    \centering
    \includegraphics[width=1.0\textwidth]{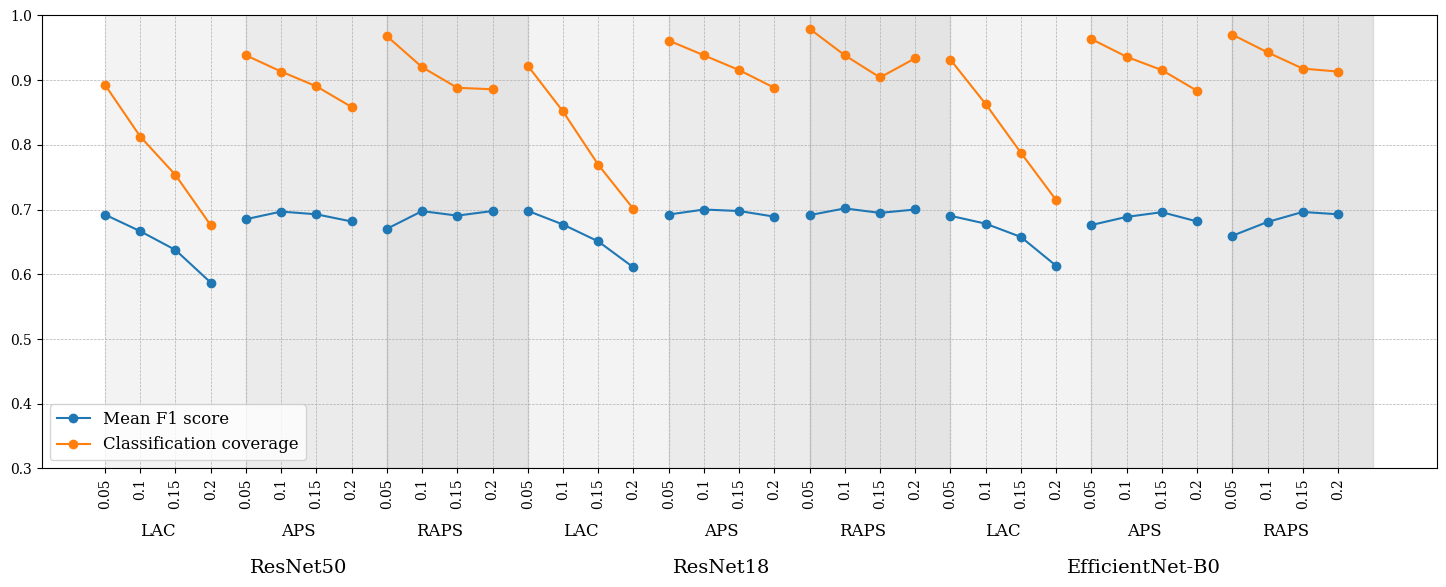}
    \caption{A summary of classification coverages and mean F1 scores of all conformal prediction methods for $\alpha \in \{0.05, 0.1, 0.15$, $0.2\}$.}
    \label{figure:f1_score_vs_coverage}
\end{figure}

Here, we report on the prediction sets' validation results using the metrics CC, SSC, mean width, mean precision, mean recall, mean F1 score, and mean Jaccard coefficient. \Cref{table:cp_classification_performance_alpha_005,table:cp_classification_performance_alpha_010,table:cp_classification_performance_alpha_015,table:cp_classification_performance_alpha_020} present the validation results with different $\alpha$ values and \Cref{figure:f1_score_vs_coverage} presents a summary of the comparison between CC and mean F1 score. CC exceeds the mean F1 score across all $\alpha$s, conformal prediction methods, and models. A value of $\alpha=0.05$ results in the highest coverage and SSC. It similarly led to the highest mean recall. The set size decreases as we increase $\alpha$, leading to high mean precision and perturbing the mean F1 score and Jaccard coefficient. Mean precision can also sometimes be higher than classification coverage (seen in parts of \Cref{table:cp_classification_performance_alpha_010,table:cp_classification_performance_alpha_015,table:cp_classification_performance_alpha_020}) because the probability of elements of the prediction sets overlapping with expert annotation sets is greater. In contrast, the classification coverage is computed against expert consensus, which contains only a single label per tile, resulting in a lower likelihood of an overlap. Increased $\alpha$ value comes at the price of losing CC and SSC. This effect is more pronounced for the LAC approach, as seen in \Cref{figure:f1_score_vs_coverage}. It also leads to cases of zero SSC in the LAC approach because, as $\alpha$ increases, LAC can result in empty prediction sets.

\begin{table}[!h]
\centering
\setlength{\extrarowheight}{0pt}
\addtolength{\extrarowheight}{\aboverulesep}
\addtolength{\extrarowheight}{\belowrulesep}
\setlength{\aboverulesep}{0pt}
\setlength{\belowrulesep}{0pt}
\caption{Performance of conformal prediction sets generated at $\alpha = 0.05$. A reverse Viridis colour scale is used per column, with yellow and dark purple highlighting the highest and the smallest values in a column, respectively.}
\label{table:cp_classification_performance_alpha_005}

\resizebox{1\linewidth}{!}{

\begin{tabular}{lllllllll} 
\toprule
Method                & Model        & CC                                                           & SSC                                                          & \begin{tabular}[c]{@{}l@{}}Mean \\width\end{tabular}         & \begin{tabular}[c]{@{}l@{}}Mean \\precision\end{tabular}     & \begin{tabular}[c]{@{}l@{}}Mean \\recall\end{tabular}        & \begin{tabular}[c]{@{}l@{}}Mean \\F1 score\end{tabular}      & \begin{tabular}[c]{@{}l@{}}Mean Jaccard \\coefficient\end{tabular}  \\ 
\midrule
\multirow{3}{*}{LAC}  & ResNet50     & {\cellcolor[rgb]{0.267,0.004,0.329}}\textcolor{white}{ 0.89} & {\cellcolor[rgb]{0.267,0.004,0.329}}\textcolor{white}{ 0.83} & {\cellcolor[rgb]{0.267,0.004,0.329}}\textcolor{white}{ 1.60} & {\cellcolor[rgb]{0.992,0.906,0.141}}0.82                     & {\cellcolor[rgb]{0.267,0.004,0.329}}\textcolor{white}{ 0.68} & {\cellcolor[rgb]{0.592,0.847,0.243}}\textcolor{white}{ 0.69} & {\cellcolor[rgb]{0.545,0.835,0.275}}\textcolor{white}{ 0.57}        \\
                      & ResNet18     & {\cellcolor[rgb]{0.184,0.412,0.553}}\textcolor{white}{ 0.92} & {\cellcolor[rgb]{0.251,0.263,0.529}}\textcolor{white}{ 0.85} & {\cellcolor[rgb]{0.278,0.145,0.459}}\textcolor{white}{ 1.69} & {\cellcolor[rgb]{0.718,0.867,0.161}}0.80                     & {\cellcolor[rgb]{0.251,0.263,0.529}}\textcolor{white}{ 0.70} & {\cellcolor[rgb]{0.992,0.906,0.141}}0.70                     & {\cellcolor[rgb]{0.992,0.906,0.141}}0.58                            \\
                      & EfficientNet-B0 & {\cellcolor[rgb]{0.145,0.514,0.553}}\textcolor{white}{ 0.93} & {\cellcolor[rgb]{0.137,0.537,0.553}}\textcolor{white}{ 0.87} & {\cellcolor[rgb]{0.212,0.353,0.549}}\textcolor{white}{ 1.85} & {\cellcolor[rgb]{0.341,0.776,0.396}}\textcolor{white}{ 0.77} & {\cellcolor[rgb]{0.188,0.404,0.553}}\textcolor{white}{ 0.72} & {\cellcolor[rgb]{0.475,0.82,0.318}}\textcolor{white}{ 0.69}  & {\cellcolor[rgb]{0.525,0.831,0.286}}\textcolor{white}{ 0.57}        \\ 
\hline
\multirow{3}{*}{APS}  & ResNet50     & {\cellcolor[rgb]{0.122,0.588,0.545}}\textcolor{white}{ 0.94} & {\cellcolor[rgb]{0.122,0.588,0.545}}\textcolor{white}{ 0.88} & {\cellcolor[rgb]{0.118,0.616,0.533}}\textcolor{white}{ 2.08} & {\cellcolor[rgb]{0.125,0.569,0.549}}\textcolor{white}{ 0.73} & {\cellcolor[rgb]{0.184,0.702,0.482}}\textcolor{white}{ 0.75} & {\cellcolor[rgb]{0.208,0.718,0.471}}\textcolor{white}{ 0.69} & {\cellcolor[rgb]{0.184,0.702,0.482}}\textcolor{white}{ 0.56}        \\
                      & ResNet18     & {\cellcolor[rgb]{0.455,0.816,0.329}}\textcolor{white}{ 0.96} & {\cellcolor[rgb]{0.129,0.651,0.522}}\textcolor{white}{ 0.89} & {\cellcolor[rgb]{0.118,0.612,0.537}}\textcolor{white}{ 2.08} & {\cellcolor[rgb]{0.125,0.561,0.549}}\textcolor{white}{ 0.72} & {\cellcolor[rgb]{0.318,0.769,0.408}}\textcolor{white}{ 0.76} & {\cellcolor[rgb]{0.604,0.847,0.235}}\textcolor{white}{ 0.69} & {\cellcolor[rgb]{0.553,0.839,0.267}}\textcolor{white}{ 0.57}        \\
                      & EfficientNet-B0 & {\cellcolor[rgb]{0.514,0.827,0.294}}\textcolor{white}{ 0.96} & {\cellcolor[rgb]{0.118,0.592,0.541}}\textcolor{white}{ 0.88} & {\cellcolor[rgb]{0.412,0.8,0.357}}\textcolor{white}{ 2.27}   & {\cellcolor[rgb]{0.208,0.365,0.549}}\textcolor{white}{ 0.68} & {\cellcolor[rgb]{0.514,0.827,0.294}}\textcolor{white}{ 0.77} & {\cellcolor[rgb]{0.149,0.498,0.557}}\textcolor{white}{ 0.68} & {\cellcolor[rgb]{0.153,0.486,0.557}}\textcolor{white}{ 0.55}        \\ 
\hline
\multirow{3}{*}{RAPS} & ResNet50     & {\cellcolor[rgb]{0.655,0.859,0.2}}\textcolor{white}{ 0.97}   & {\cellcolor[rgb]{0.169,0.451,0.557}}\textcolor{white}{ 0.86} & {\cellcolor[rgb]{0.447,0.812,0.333}}\textcolor{white}{ 2.29} & {\cellcolor[rgb]{0.239,0.29,0.537}}\textcolor{white}{ 0.67}  & {\cellcolor[rgb]{0.647,0.855,0.208}}\textcolor{white}{ 0.78} & {\cellcolor[rgb]{0.216,0.349,0.549}}\textcolor{white}{ 0.67} & {\cellcolor[rgb]{0.243,0.286,0.537}}\textcolor{white}{ 0.54}        \\
                      & ResNet18     & {\cellcolor[rgb]{0.992,0.906,0.141}}0.98                     & {\cellcolor[rgb]{0.992,0.906,0.141}}0.93                     & {\cellcolor[rgb]{0.251,0.741,0.447}}\textcolor{white}{ 2.21} & {\cellcolor[rgb]{0.18,0.42,0.557}}\textcolor{white}{ 0.69}   & {\cellcolor[rgb]{0.78,0.878,0.122}}\textcolor{white}{ 0.78}  & {\cellcolor[rgb]{0.545,0.835,0.275}}\textcolor{white}{ 0.69} & {\cellcolor[rgb]{0.447,0.812,0.333}}\textcolor{white}{ 0.57}        \\
                      & EfficientNet-B0 & {\cellcolor[rgb]{0.729,0.871,0.153}}\textcolor{white}{ 0.97} & {\cellcolor[rgb]{0.259,0.745,0.443}}\textcolor{white}{ 0.90} & {\cellcolor[rgb]{0.992,0.906,0.141}}2.47                     & {\cellcolor[rgb]{0.267,0.004,0.329}}\textcolor{white}{ 0.63} & {\cellcolor[rgb]{0.992,0.906,0.141}}0.79                     & {\cellcolor[rgb]{0.267,0.004,0.329}}\textcolor{white}{ 0.66} & {\cellcolor[rgb]{0.267,0.004,0.329}}\textcolor{white}{ 0.53}        \\
\bottomrule
\end{tabular}
} 
\end{table}

\begin{table}[!h]
\centering
\setlength{\extrarowheight}{0pt}
\addtolength{\extrarowheight}{\aboverulesep}
\addtolength{\extrarowheight}{\belowrulesep}
\setlength{\aboverulesep}{0pt}
\setlength{\belowrulesep}{0pt}
\caption{Performance of conformal prediction sets generated at $\alpha = 0.1$. A reverse Viridis colour scale is used per column, with yellow and dark purple highlighting the highest and the smallest values in a column, respectively.}
\label{table:cp_classification_performance_alpha_010}
\resizebox{1\linewidth}{!}{%

\begin{tabular}{lllllllll} 
\toprule
Method                & Model        & CC                                                           & SSC                                                          & \begin{tabular}[c]{@{}l@{}}Mean \\width\end{tabular}         & \begin{tabular}[c]{@{}l@{}}Mean\\precision\end{tabular}      & \begin{tabular}[c]{@{}l@{}}Mean \\recall\end{tabular}        & \begin{tabular}[c]{@{}l@{}}Mean \\F1 score\end{tabular}      & \begin{tabular}[c]{@{}l@{}}Mean Jaccard\\coefficient\end{tabular}  \\ 
\midrule
\multirow{3}{*}{LAC}  & ResNet50     & {\cellcolor[rgb]{0.267,0.004,0.329}}\textcolor{white}{ 0.81} & {\cellcolor[rgb]{0.267,0.004,0.329}}\textcolor{white}{ 0.78} & {\cellcolor[rgb]{0.267,0.004,0.329}}\textcolor{white}{ 1.25} & {\cellcolor[rgb]{0.992,0.906,0.141}}0.86                     & {\cellcolor[rgb]{0.267,0.004,0.329}}\textcolor{white}{ 0.60} & {\cellcolor[rgb]{0.267,0.004,0.329}}\textcolor{white}{ 0.67} & {\cellcolor[rgb]{0.267,0.004,0.329}}\textcolor{white}{ 0.55}       \\
                      & ResNet18     & {\cellcolor[rgb]{0.204,0.369,0.553}}\textcolor{white}{ 0.85} & {\cellcolor[rgb]{0.239,0.294,0.537}}\textcolor{white}{ 0.80} & {\cellcolor[rgb]{0.278,0.145,0.459}}\textcolor{white}{ 1.33} & {\cellcolor[rgb]{0.831,0.882,0.102}}\textcolor{white}{ 0.86} & {\cellcolor[rgb]{0.271,0.212,0.506}}\textcolor{white}{ 0.62} & {\cellcolor[rgb]{0.208,0.365,0.549}}\textcolor{white}{ 0.68} & {\cellcolor[rgb]{0.184,0.412,0.553}}\textcolor{white}{ 0.56}       \\
                      & EfficientNet-B0 & {\cellcolor[rgb]{0.169,0.455,0.557}}\textcolor{white}{ 0.86} & {\cellcolor[rgb]{0.149,0.502,0.557}}\textcolor{white}{ 0.82} & {\cellcolor[rgb]{0.22,0.337,0.545}}\textcolor{white}{ 1.46}  & {\cellcolor[rgb]{0.392,0.796,0.365}}\textcolor{white}{ 0.83} & {\cellcolor[rgb]{0.204,0.373,0.553}}\textcolor{white}{ 0.64} & {\cellcolor[rgb]{0.192,0.4,0.553}}\textcolor{white}{ 0.68}   & {\cellcolor[rgb]{0.18,0.42,0.557}}\textcolor{white}{ 0.56}         \\ 
\hline
\multirow{3}{*}{APS}  & ResNet50     & {\cellcolor[rgb]{0.412,0.8,0.357}}\textcolor{white}{ 0.91}   & {\cellcolor[rgb]{0.678,0.863,0.188}}\textcolor{white}{ 0.86} & {\cellcolor[rgb]{0.129,0.651,0.522}}\textcolor{white}{ 1.71} & {\cellcolor[rgb]{0.118,0.6,0.541}}\textcolor{white}{ 0.80}   & {\cellcolor[rgb]{0.376,0.788,0.376}}\textcolor{white}{ 0.71} & {\cellcolor[rgb]{0.635,0.855,0.216}}\textcolor{white}{ 0.70} & {\cellcolor[rgb]{0.439,0.808,0.337}}\textcolor{white}{ 0.58}       \\
                      & ResNet18     & {\cellcolor[rgb]{0.914,0.894,0.098}}0.94                     & {\cellcolor[rgb]{0.573,0.843,0.255}}\textcolor{white}{ 0.86} & {\cellcolor[rgb]{0.475,0.82,0.318}}\textcolor{white}{ 1.87}  & {\cellcolor[rgb]{0.208,0.361,0.549}}\textcolor{white}{ 0.77} & {\cellcolor[rgb]{0.933,0.898,0.106}}0.74                     & {\cellcolor[rgb]{0.875,0.89,0.094}}\textcolor{white}{ 0.70}  & {\cellcolor[rgb]{0.729,0.871,0.153}}\textcolor{white}{ 0.58}       \\
                      & EfficientNet-B0 & {\cellcolor[rgb]{0.863,0.886,0.094}}0.94                     & {\cellcolor[rgb]{0.741,0.871,0.149}}\textcolor{white}{ 0.86} & {\cellcolor[rgb]{0.678,0.863,0.188}}\textcolor{white}{ 1.93} & {\cellcolor[rgb]{0.259,0.239,0.518}}\textcolor{white}{ 0.75} & {\cellcolor[rgb]{0.831,0.882,0.102}}0.73                     & {\cellcolor[rgb]{0.157,0.682,0.498}}\textcolor{white}{ 0.69} & {\cellcolor[rgb]{0.129,0.557,0.549}}\textcolor{white}{ 0.57}       \\ 
\hline
\multirow{3}{*}{RAPS} & ResNet50     & {\cellcolor[rgb]{0.545,0.835,0.275}}\textcolor{white}{ 0.92} & {\cellcolor[rgb]{0.992,0.906,0.141}}0.87                     & {\cellcolor[rgb]{0.137,0.659,0.514}}\textcolor{white}{ 1.71} & {\cellcolor[rgb]{0.122,0.584,0.545}}\textcolor{white}{ 0.80} & {\cellcolor[rgb]{0.412,0.8,0.357}}\textcolor{white}{ 0.71}   & {\cellcolor[rgb]{0.686,0.863,0.18}}\textcolor{white}{ 0.70}  & {\cellcolor[rgb]{0.475,0.82,0.318}}\textcolor{white}{ 0.58}        \\
                      & ResNet18     & {\cellcolor[rgb]{0.914,0.894,0.098}}0.94                     & {\cellcolor[rgb]{0.655,0.859,0.2}}\textcolor{white}{ 0.86}   & {\cellcolor[rgb]{0.412,0.8,0.357}}\textcolor{white}{ 1.84}   & {\cellcolor[rgb]{0.196,0.388,0.553}}\textcolor{white}{ 0.77} & {\cellcolor[rgb]{0.894,0.89,0.094}}0.73                      & {\cellcolor[rgb]{0.992,0.906,0.141}}0.70                     & {\cellcolor[rgb]{0.992,0.906,0.141}}0.58                           \\
                      & EfficientNet-B0 & {\cellcolor[rgb]{0.992,0.906,0.141}}0.94                     & {\cellcolor[rgb]{0.667,0.859,0.196}}\textcolor{white}{ 0.86} & {\cellcolor[rgb]{0.992,0.906,0.141}}2.02                     & {\cellcolor[rgb]{0.267,0.004,0.329}}\textcolor{white}{ 0.73} & {\cellcolor[rgb]{0.992,0.906,0.141}}0.74                     & {\cellcolor[rgb]{0.161,0.475,0.557}}\textcolor{white}{ 0.68} & {\cellcolor[rgb]{0.271,0.208,0.502}}\textcolor{white}{ 0.55}       \\
\bottomrule
\end{tabular}
} 

\end{table}

\clearpage  

\begin{table}[!h]
\centering
\setlength{\extrarowheight}{0pt}
\addtolength{\extrarowheight}{\aboverulesep}
\addtolength{\extrarowheight}{\belowrulesep}
\setlength{\aboverulesep}{0pt}
\setlength{\belowrulesep}{0pt}
\caption{Performance of conformal prediction sets generated at $\alpha = 0.15$. A reverse Viridis colour scale is used per column, with yellow and dark purple highlighting the highest and the smallest values in a column, respectively.}
\label{table:cp_classification_performance_alpha_015}
\resizebox{1\linewidth}{!}{
\begin{tabular}{lllllllll} 
\toprule
Method                & Model        & CC                                                           & SSC                                                          & \begin{tabular}[c]{@{}l@{}}Mean \\width\end{tabular}         & \begin{tabular}[c]{@{}l@{}}Mean\\precision\end{tabular}      & \begin{tabular}[c]{@{}l@{}}Mean \\recall\end{tabular}        & \begin{tabular}[c]{@{}l@{}}Mean \\F1 score\end{tabular}      & \begin{tabular}[c]{@{}l@{}}Mean Jaccard\\coefficient\end{tabular}  \\ 
\midrule
\multirow{3}{*}{LAC}  & ResNet50     & {\cellcolor[rgb]{0.267,0.004,0.329}}\textcolor{white}{ 0.75} & {\cellcolor[rgb]{0.267,0.004,0.329}}\textcolor{white}{ 0.00} & {\cellcolor[rgb]{0.267,0.004,0.329}}\textcolor{white}{ 1.06} & {\cellcolor[rgb]{0.855,0.886,0.094}}0.88                     & {\cellcolor[rgb]{0.267,0.004,0.329}}\textcolor{white}{ 0.54} & {\cellcolor[rgb]{0.267,0.004,0.329}}\textcolor{white}{ 0.64} & {\cellcolor[rgb]{0.267,0.004,0.329}}\textcolor{white}{ 0.52}  \\
                      & ResNet18     & {\cellcolor[rgb]{0.282,0.133,0.451}}\textcolor{white}{ 0.77} & {\cellcolor[rgb]{0.584,0.843,0.247}}\textcolor{white}{ 0.74} & {\cellcolor[rgb]{0.278,0.173,0.482}}\textcolor{white}{ 1.14} & {\cellcolor[rgb]{0.71,0.867,0.169}}\textcolor{white}{ 0.87}  & {\cellcolor[rgb]{0.275,0.176,0.486}}\textcolor{white}{ 0.56} & {\cellcolor[rgb]{0.243,0.286,0.537}}\textcolor{white}{ 0.65} & {\cellcolor[rgb]{0.22,0.341,0.549}}\textcolor{white}{ 0.54}   \\
                      & EfficientNet-B0 & {\cellcolor[rgb]{0.247,0.271,0.529}}\textcolor{white}{ 0.79} & {\cellcolor[rgb]{0.686,0.863,0.18}}\textcolor{white}{ 0.77}  & {\cellcolor[rgb]{0.278,0.173,0.482}}\textcolor{white}{ 1.14} & {\cellcolor[rgb]{0.992,0.906,0.141}}0.88                     & {\cellcolor[rgb]{0.263,0.227,0.514}}\textcolor{white}{ 0.57} & {\cellcolor[rgb]{0.188,0.404,0.553}}\textcolor{white}{ 0.66} & {\cellcolor[rgb]{0.161,0.471,0.557}}\textcolor{white}{ 0.54}  \\ 
\hline
\multirow{3}{*}{APS}  & ResNet50     & {\cellcolor[rgb]{0.565,0.839,0.263}}\textcolor{white}{ 0.89} & {\cellcolor[rgb]{0.855,0.886,0.094}}0.83                     & {\cellcolor[rgb]{0.467,0.816,0.322}}\textcolor{white}{ 1.57} & {\cellcolor[rgb]{0.188,0.404,0.553}}\textcolor{white}{ 0.83} & {\cellcolor[rgb]{0.624,0.851,0.22}}\textcolor{white}{ 0.68}  & {\cellcolor[rgb]{0.78,0.878,0.122}}0.69                      & {\cellcolor[rgb]{0.616,0.851,0.227}}\textcolor{white}{ 0.57}  \\
                      & ResNet18     & {\cellcolor[rgb]{0.965,0.902,0.122}}0.92                     & {\cellcolor[rgb]{0.906,0.894,0.098}}0.84                     & {\cellcolor[rgb]{0.875,0.89,0.094}}1.67                      & {\cellcolor[rgb]{0.278,0.078,0.4}}\textcolor{white}{ 0.80}   & {\cellcolor[rgb]{0.98,0.902,0.133}}0.70                      & {\cellcolor[rgb]{0.992,0.906,0.141}}0.70                     & {\cellcolor[rgb]{0.992,0.906,0.141}}0.58                      \\
                      & EfficientNet-B0 & {\cellcolor[rgb]{0.965,0.902,0.122}}0.92                     & {\cellcolor[rgb]{0.965,0.902,0.122}}0.86                     & {\cellcolor[rgb]{0.953,0.898,0.118}}1.69                     & {\cellcolor[rgb]{0.275,0.055,0.38}}\textcolor{white}{ 0.80}  & {\cellcolor[rgb]{0.965,0.902,0.122}}0.70                     & {\cellcolor[rgb]{0.925,0.894,0.102}}0.70                     & {\cellcolor[rgb]{0.894,0.89,0.094}}0.58                       \\ 
\hline
\multirow{3}{*}{RAPS} & ResNet50     & {\cellcolor[rgb]{0.525,0.831,0.286}}\textcolor{white}{ 0.89} & {\cellcolor[rgb]{0.831,0.882,0.102}}0.82                     & {\cellcolor[rgb]{0.439,0.808,0.337}}\textcolor{white}{ 1.56} & {\cellcolor[rgb]{0.192,0.4,0.553}}\textcolor{white}{ 0.83}   & {\cellcolor[rgb]{0.565,0.839,0.263}}\textcolor{white}{ 0.67} & {\cellcolor[rgb]{0.686,0.863,0.18}}0.69                      & {\cellcolor[rgb]{0.565,0.839,0.263}}\textcolor{white}{ 0.57}  \\
                      & ResNet18     & {\cellcolor[rgb]{0.78,0.878,0.122}}0.90                      & {\cellcolor[rgb]{0.875,0.89,0.094}}0.84                      & {\cellcolor[rgb]{0.647,0.855,0.208}}\textcolor{white}{ 1.61} & {\cellcolor[rgb]{0.271,0.196,0.498}}\textcolor{white}{ 0.81} & {\cellcolor[rgb]{0.78,0.878,0.122}}0.69                      & {\cellcolor[rgb]{0.875,0.89,0.094}}0.69                      & {\cellcolor[rgb]{0.906,0.894,0.098}}0.58                      \\
                      & EfficientNet-B0 & {\cellcolor[rgb]{0.992,0.906,0.141}}0.92                     & {\cellcolor[rgb]{0.992,0.906,0.141}}0.88                     & {\cellcolor[rgb]{0.992,0.906,0.141}}1.70                     & {\cellcolor[rgb]{0.267,0.004,0.329}}\textcolor{white}{ 0.80} & {\cellcolor[rgb]{0.992,0.906,0.141}}0.70                     & {\cellcolor[rgb]{0.933,0.898,0.106}}0.70                     & {\cellcolor[rgb]{0.965,0.902,0.122}}0.58                      \\
\bottomrule
\end{tabular}
} 
\end{table}

\begin{table}[!h]
\centering
\setlength{\extrarowheight}{0pt}
\addtolength{\extrarowheight}{\aboverulesep}
\addtolength{\extrarowheight}{\belowrulesep}
\setlength{\aboverulesep}{0pt}
\setlength{\belowrulesep}{0pt}
\caption{Performance of conformal prediction sets generated at $\alpha = 0.2$.  A reverse Viridis colour scale is used per column, with yellow and dark purple highlighting the highest and the smallest values in a column, respectively.}
\label{table:cp_classification_performance_alpha_020}
\resizebox{1\linewidth}{!}{%

\begin{tabular}{lllllllll} 
\toprule
Method                & Model        & CC                                                           & SSC                                                          & \begin{tabular}[c]{@{}l@{}}Mean \\width\end{tabular}         & \begin{tabular}[c]{@{}l@{}}Mean\\precision\end{tabular}      & \begin{tabular}[c]{@{}l@{}}Mean \\recall\end{tabular}        & \begin{tabular}[c]{@{}l@{}}Mean \\F1 score\end{tabular}      & \begin{tabular}[c]{@{}l@{}}Mean Jaccard\\coefficient\end{tabular}  \\ 
\midrule
\multirow{3}{*}{LAC}  & ResNet50     & {\cellcolor[rgb]{0.267,0.004,0.329}}\textcolor{white}{ 0.68} & {\cellcolor[rgb]{0.267,0.004,0.329}}\textcolor{white}{ 0.00} & {\cellcolor[rgb]{0.267,0.004,0.329}}\textcolor{white}{ 0.92} & {\cellcolor[rgb]{0.404,0.8,0.361}}\textcolor{white}{ 0.83}   & {\cellcolor[rgb]{0.267,0.004,0.329}}\textcolor{white}{ 0.48} & {\cellcolor[rgb]{0.267,0.004,0.329}}\textcolor{white}{ 0.59} & {\cellcolor[rgb]{0.267,0.004,0.329}}\textcolor{white}{ 0.48}  \\
                      & ResNet18     & {\cellcolor[rgb]{0.282,0.133,0.451}}\textcolor{white}{ 0.70} & {\cellcolor[rgb]{0.267,0.004,0.329}}\textcolor{white}{ 0.00} & {\cellcolor[rgb]{0.275,0.043,0.369}}\textcolor{white}{ 0.95} & {\cellcolor[rgb]{0.992,0.906,0.141}}0.86                     & {\cellcolor[rgb]{0.282,0.125,0.443}}\textcolor{white}{ 0.50} & {\cellcolor[rgb]{0.247,0.271,0.529}}\textcolor{white}{ 0.61} & {\cellcolor[rgb]{0.239,0.29,0.537}}\textcolor{white}{ 0.50}   \\
                      & EfficientNet-B0 & {\cellcolor[rgb]{0.271,0.204,0.498}}\textcolor{white}{ 0.71} & {\cellcolor[rgb]{0.267,0.004,0.329}}\textcolor{white}{ 0.00} & {\cellcolor[rgb]{0.275,0.047,0.373}}\textcolor{white}{ 0.95} & {\cellcolor[rgb]{0.992,0.906,0.141}}0.86                     & {\cellcolor[rgb]{0.278,0.145,0.459}}\textcolor{white}{ 0.51} & {\cellcolor[rgb]{0.239,0.294,0.537}}\textcolor{white}{ 0.61} & {\cellcolor[rgb]{0.224,0.333,0.545}}\textcolor{white}{ 0.51}  \\ 
\hline
\multirow{3}{*}{APS}  & ResNet50     & {\cellcolor[rgb]{0.278,0.753,0.431}}\textcolor{white}{ 0.86} & {\cellcolor[rgb]{0.749,0.875,0.141}}0.81                     & {\cellcolor[rgb]{0.122,0.635,0.525}}\textcolor{white}{ 1.44} & {\cellcolor[rgb]{0.678,0.863,0.188}}\textcolor{white}{ 0.84} & {\cellcolor[rgb]{0.161,0.686,0.498}}\textcolor{white}{ 0.64} & {\cellcolor[rgb]{0.573,0.843,0.255}}\textcolor{white}{ 0.68} & {\cellcolor[rgb]{0.486,0.824,0.31}}\textcolor{white}{ 0.56}   \\
                      & ResNet18     & {\cellcolor[rgb]{0.533,0.835,0.278}}\textcolor{white}{ 0.89} & {\cellcolor[rgb]{0.824,0.882,0.106}}0.84                     & {\cellcolor[rgb]{0.184,0.702,0.482}}\textcolor{white}{ 1.51} & {\cellcolor[rgb]{0.392,0.796,0.365}}\textcolor{white}{ 0.83} & {\cellcolor[rgb]{0.294,0.761,0.424}}\textcolor{white}{ 0.66} & {\cellcolor[rgb]{0.741,0.871,0.149}}0.69                     & {\cellcolor[rgb]{0.678,0.863,0.188}}0.57                      \\
                      & EfficientNet-B0 & {\cellcolor[rgb]{0.494,0.824,0.306}}\textcolor{white}{ 0.88} & {\cellcolor[rgb]{0.831,0.882,0.102}}0.84                     & {\cellcolor[rgb]{0.251,0.741,0.447}}\textcolor{white}{ 1.55} & {\cellcolor[rgb]{0.18,0.698,0.486}}\textcolor{white}{ 0.81}  & {\cellcolor[rgb]{0.267,0.745,0.439}}\textcolor{white}{ 0.66} & {\cellcolor[rgb]{0.573,0.843,0.255}}\textcolor{white}{ 0.68} & {\cellcolor[rgb]{0.553,0.839,0.267}}\textcolor{white}{ 0.56}  \\ 
\hline
\multirow{3}{*}{RAPS} & ResNet50     & {\cellcolor[rgb]{0.514,0.827,0.294}}\textcolor{white}{ 0.89} & {\cellcolor[rgb]{0.894,0.89,0.094}}0.86                      & {\cellcolor[rgb]{0.243,0.737,0.451}}\textcolor{white}{ 1.54} & {\cellcolor[rgb]{0.267,0.745,0.439}}\textcolor{white}{ 0.82} & {\cellcolor[rgb]{0.427,0.808,0.345}}\textcolor{white}{ 0.68} & {\cellcolor[rgb]{0.945,0.898,0.11}}0.70                      & {\cellcolor[rgb]{0.98,0.902,0.133}}0.58                       \\
                      & ResNet18     & {\cellcolor[rgb]{0.992,0.906,0.141}}0.93                     & {\cellcolor[rgb]{0.992,0.906,0.141}}0.90                     & {\cellcolor[rgb]{0.992,0.906,0.141}}1.82                     & {\cellcolor[rgb]{0.278,0.082,0.404}}\textcolor{white}{ 0.74} & {\cellcolor[rgb]{0.992,0.906,0.141}}0.74                     & {\cellcolor[rgb]{0.992,0.906,0.141}}0.70                     & {\cellcolor[rgb]{0.992,0.906,0.141}}0.58                      \\
                      & EfficientNet-B0 & {\cellcolor[rgb]{0.792,0.878,0.118}}\textcolor{white}{ 0.91} & {\cellcolor[rgb]{0.945,0.898,0.11}}0.88                      & {\cellcolor[rgb]{0.98,0.902,0.133}}1.82                      & {\cellcolor[rgb]{0.267,0.004,0.329}}\textcolor{white}{ 0.74} & {\cellcolor[rgb]{0.831,0.882,0.102}}0.72                     & {\cellcolor[rgb]{0.831,0.882,0.102}}0.69                     & {\cellcolor[rgb]{0.882,0.89,0.094}}0.58                       \\
\bottomrule
\end{tabular}
} 
\end{table}
\clearpage

\subsection{Conformal prediction sets are not well aligned with experts' annotations}
\label{section:results:pairwise_alignment_with_experts}

\begin{figure}[!h]
    \centering
    \begin{subfigure}{\textwidth}
        \centering
        \includegraphics[width=1\textwidth]{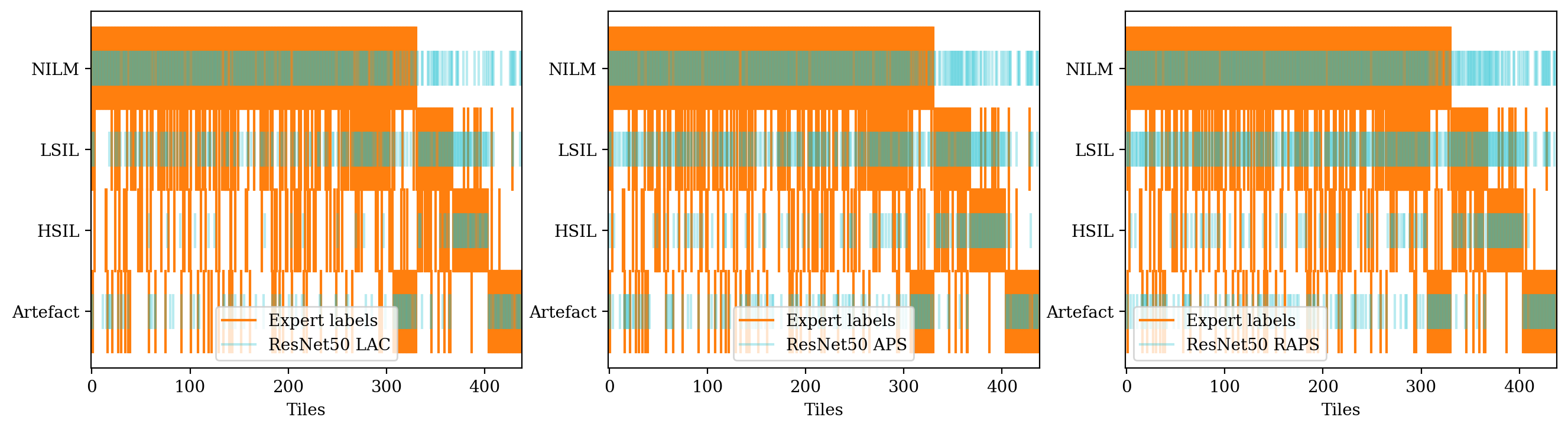}
    \end{subfigure}
    
    \begin{subfigure}{\textwidth}
        \centering
        \includegraphics[width=1\textwidth]{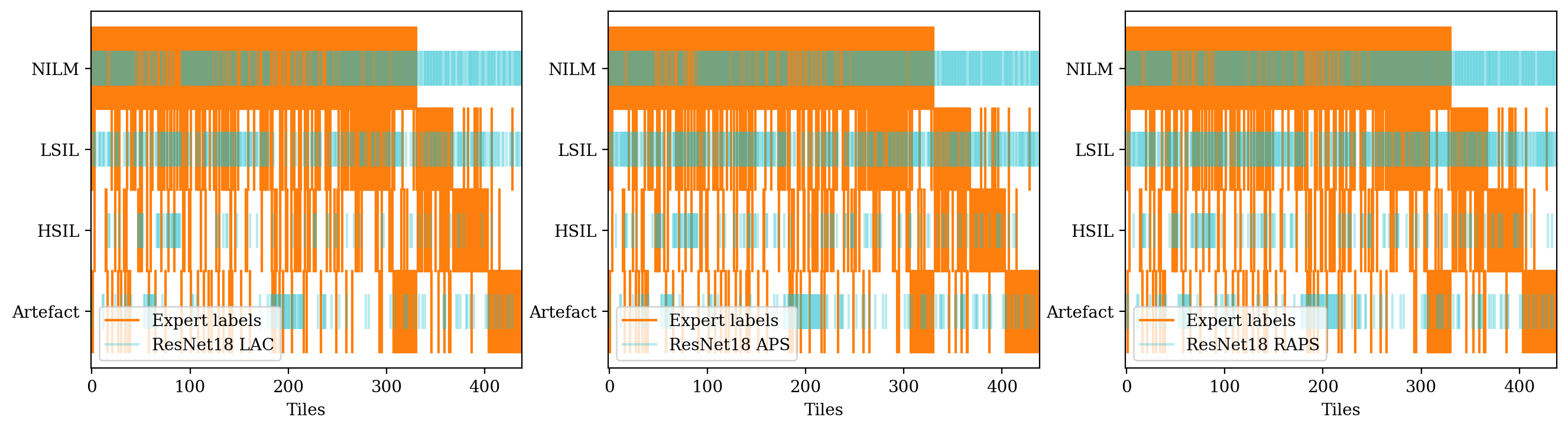}
    \end{subfigure}
    
    \begin{subfigure}{\textwidth}
        \centering
        \includegraphics[width=1\textwidth]{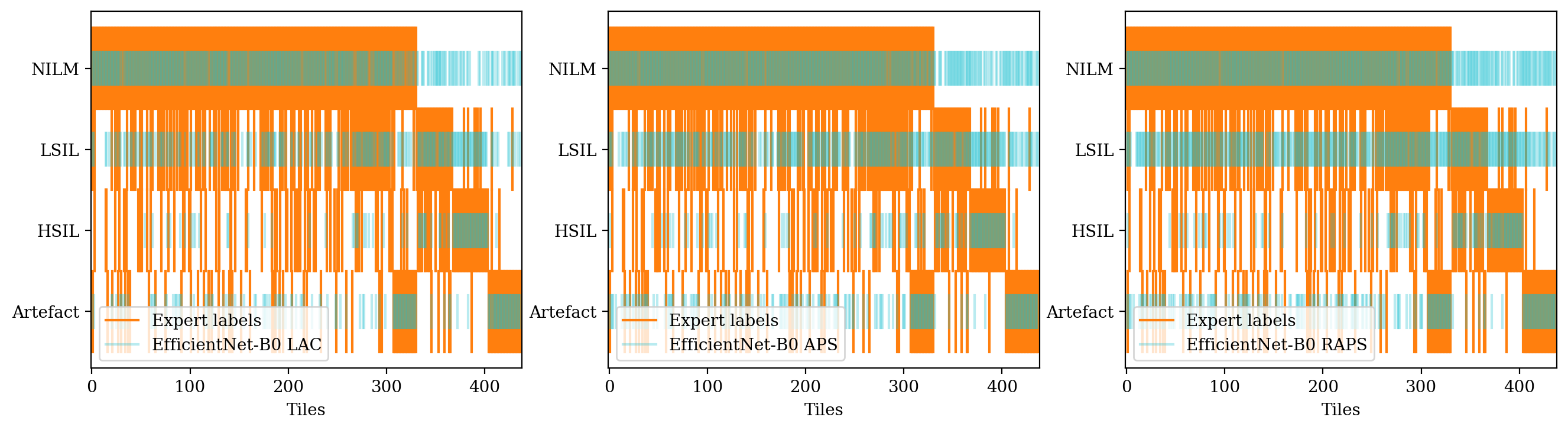}
    \end{subfigure}

    \caption{Highlight of individual ground truth categories accurately identified by conformal prediction sets ($\alpha = 0.05$). For each of the tiles, we expected the perpendicular red lines to overlap with the light blue lines across all the categories, which ended up not being the case.}
    \label{figure:set-by_set_comparison}
\end{figure}

Here, we assess if the prediction sets correctly localize all the labels given by experts for each tile. \Cref{table:set-by_set_comparison} summarizes the pairwise comparison between the experts' annotation sets and the conformal prediction sets for exact matches. This comparison is one of the rare cases where almost all the methods fail to accurately generate the sets collected from experts to an acceptable level (accuracy = 0.33 $\pm$ 0.04). This is visualized in \Cref{figure:set-by_set_comparison} for all the test tiles. Good prediction sets would overlap with the expert labels highlighted with orange in \Cref{figure:set-by_set_comparison}.

\begin{table}[!h]
    \centering
    \caption{A summary of pairwise comparison of conformal prediction sets and expert annotation sets for exact matches.}
    \label{table:set-by_set_comparison}
    \begin{tblr}{
      cell{1}{3} = {c=4}{c},
      cell{3}{1} = {r=3}{},
      cell{6}{1} = {r=3}{},
      cell{9}{1} = {r=3}{},
      hline{1} = {3-6}{},
      hline{2-3,6,9,12} = {-}{},
    }
           &              & Accuracy      &              &               &              \\
    Method & Model        & $\alpha=$0.05 & $\alpha=$0.1 & $\alpha=$0.15 & $\alpha=$0.2 \\
    LAC    & ResNet50     & 0.34          & 0.36         & 0.35          & 0.34         \\
           & ResNet18     & 0.34          & 0.36         & 0.35          & 0.35         \\
           & EfficientNet-B0 & 0.32          & 0.35         & 0.36          & 0.37         \\
    APS    & ResNet50     & 0.29          & 0.35         & 0.34          & 0.34         \\
           & ResNet18     & 0.29          & 0.33         & 0.34          & 0.34         \\
           & EfficientNet-B0 & 0.26          & 0.31         & 0.34          & 0.32         \\
    RAPS   & ResNet50     & 0.22          & 0.34         & 0.34          & 0.35         \\
           & ResNet18     & 0.27          & 0.34         & 0.35          & 0.34         \\
           & EfficientNet-B0 & 0.18          & 0.28         & 0.35          & 0.34         
    \end{tblr}
\end{table}

\subsection{Performance in capturing aleatoric uncertainty}
\label{section:results:aleatoric}


Here, we assess the conformal prediction sets' performance in correctly capturing aleatoric uncertainty. We chose an alpha of $\alpha=0.05$ for all the conformal prediction methods since this value led to the highest classification coverage and SSC across all the methods and deep learning models (See \Cref{table:cp_classification_performance_alpha_005}).  

\begin{figure}[h]
    \centering
    \begin{subfigure}{\textwidth}
        \centering
        \includegraphics[width=1\textwidth]{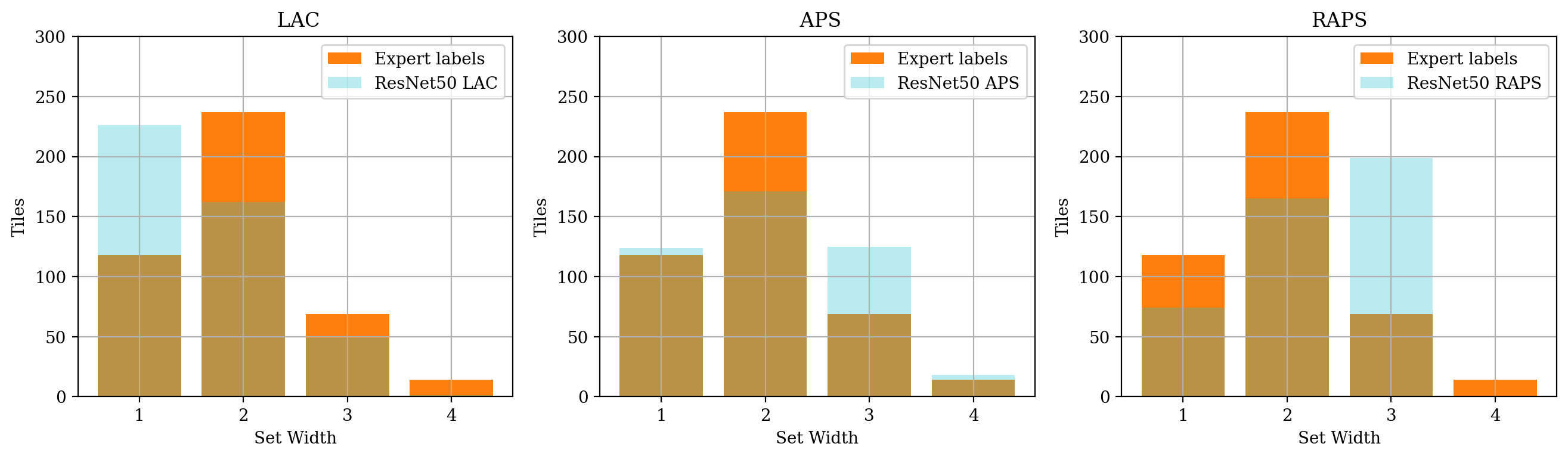}
    \end{subfigure}
    
    \begin{subfigure}{\textwidth}
        \centering
        \includegraphics[width=1\textwidth]{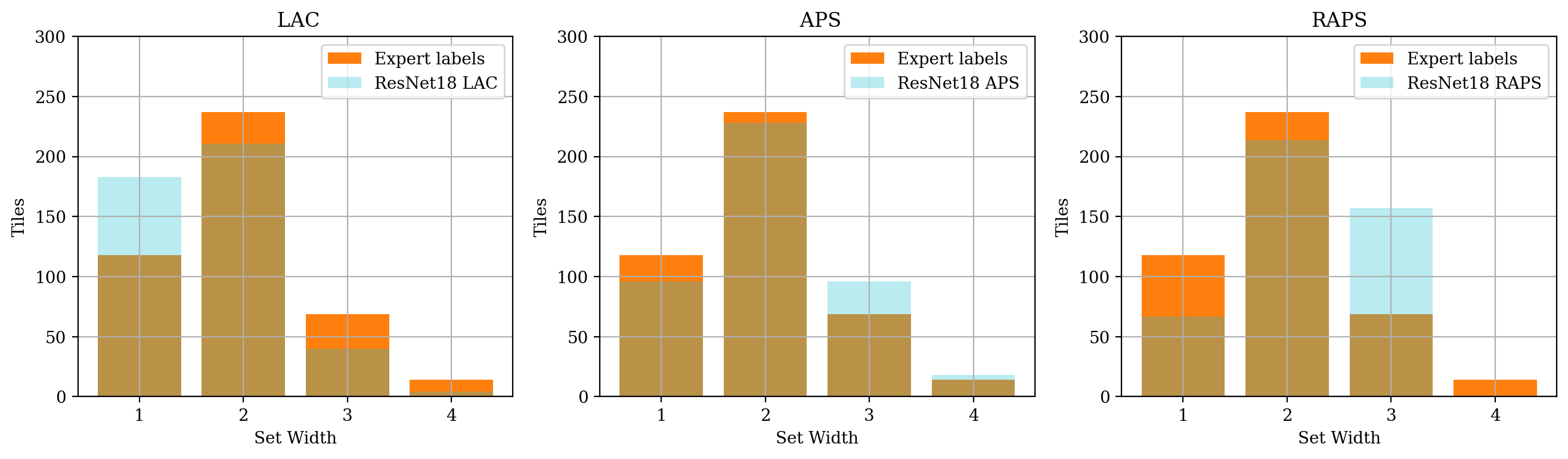}
    \end{subfigure}
    
    \begin{subfigure}{\textwidth}
        \centering
        \includegraphics[width=1\textwidth]{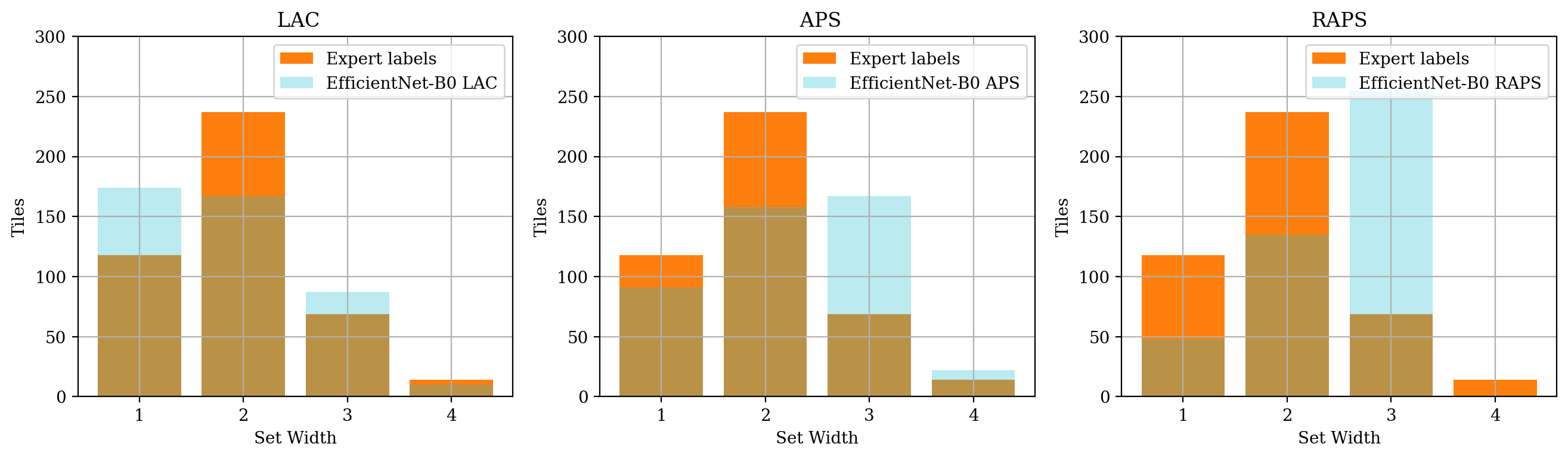}
    \end{subfigure}

    \caption{Performance of conformal prediction in capturing tile ambiguity.}
    \label{figure:capturing_aleatoric_uncertainty}
\end{figure}

The conformal prediction methods' performance in capturing aleatoric uncertainty for all three models is plotted in \Cref{figure:capturing_aleatoric_uncertainty} with a summary presented in \Cref{table:tile_ambiguity_capturing_performance}. APS achieves the highest performance of 92.92\%.


\begin{table}[!h]
\centering
\caption{Performance of conformal prediction sets ($\alpha = 0.05$) in capturing tile ambiguity.}
\label{table:tile_ambiguity_capturing_performance}
\begin{tblr}{
  vline{2} = {-}{},
  hline{1-2,5} = {-}{},
}
Conformal prediction methods      & LAC   & APS    & RAPS  \\
ResNet50        & 75.34\% & 84.93\%  & 70.32\% \\
ResNet18        & 85.16\% & 92.92\%  & 79.91\% \\
EfficientNet-B0 & 83.11\% & 75.79\%  & 57.53\% 
\end{tblr}
\end{table}


\subsection{Performance in capturing epistemic uncertainty}
\label{section:results:epistemic}


Here, we report on the performance of all conformal prediction methods in detecting the two OOD datasets. As in the previous section, an alpha value of $\alpha=0.05$ is used for all the conformal prediction methods. 

\subsubsection{First OOD: noised test data}
\label{section:results_ood_detction_noised_data}

\begin{figure}[h]
    \centering
    \begin{subfigure}{\textwidth}
        \centering
        \includegraphics[width=1\textwidth]{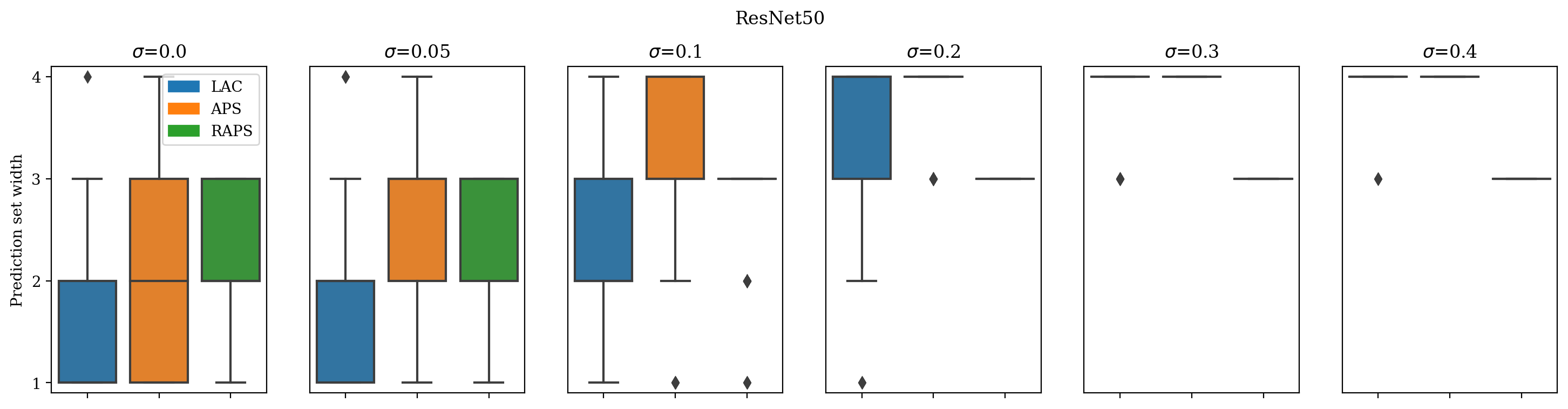}
    \end{subfigure}
    
    \begin{subfigure}{\textwidth}
        \centering
        \includegraphics[width=1\textwidth]{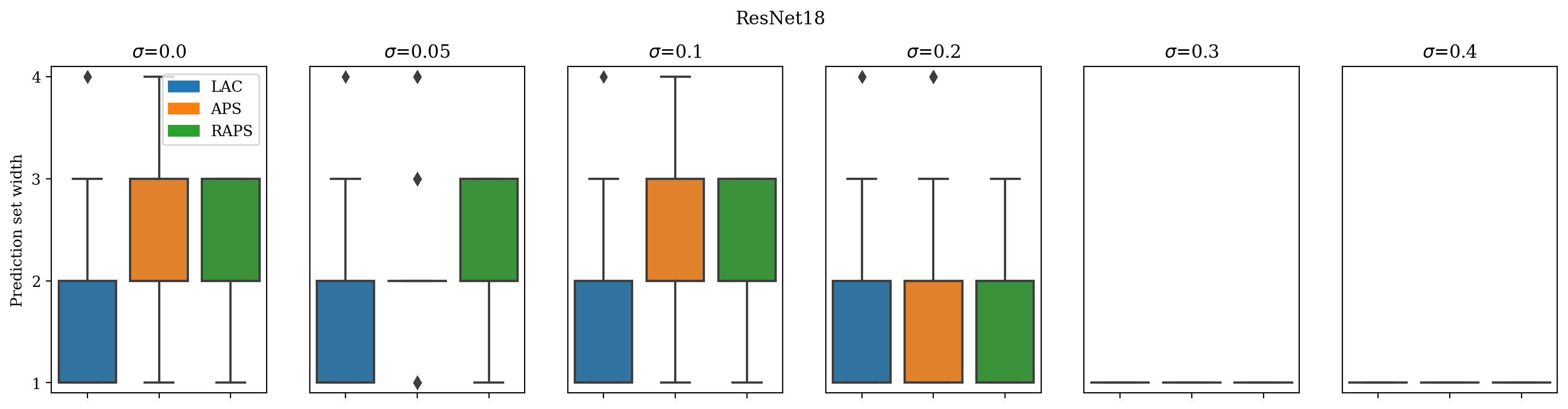}
    \end{subfigure}
    
    \begin{subfigure}{\textwidth}
        \centering
        \includegraphics[width=1\textwidth]{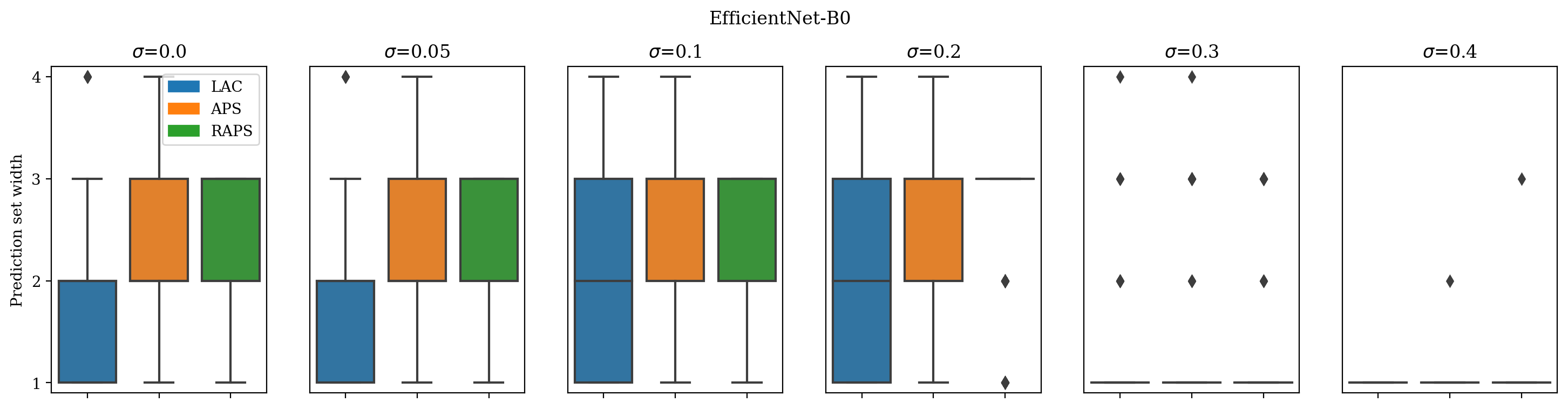}
    \end{subfigure}

    \caption{Capability of Conformal prediction methods in detecting Out-of-Distribution (OOD) data generated by adding noise to the test data. We expect the set width to increase relative to the noise level. However, the trend shows a model-dependent OOD detection where only the prediction set width of the ResNet50 model shows an increase.}
    \label{figure:ood_detection_results}
\end{figure}

\Cref{figure:ood_detection_results} summarizes the performance of conformal prediction in detecting OOD versions of our test set with sequentially added noise. The conformal prediction methods applied to the ResNet50 model show increased set width. However, they do not work reflect on the ResNet18 and EfficientNet-B0 models. The trend in \Cref{figure:ood_detection_results} shows that the conformal prediction methods' capability in detecting the noised OOD data is model-dependent.


\subsubsection{Second OOD: bone marrow cytomorphology data}

\begin{figure}[h]
    \centering
    \includegraphics[width=1.0\textwidth]{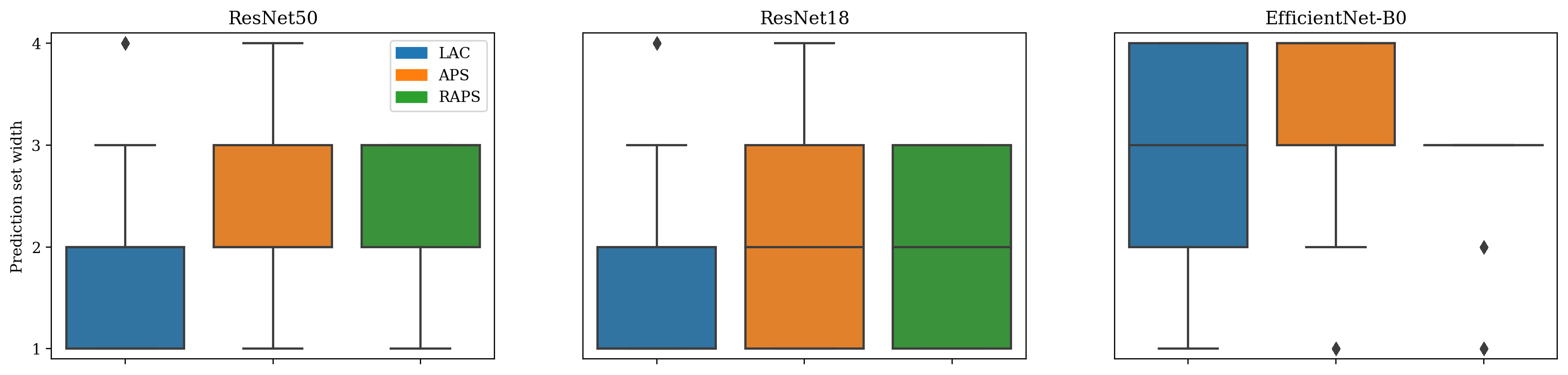}
    \caption{Performance of Conformal prediction methods in detecting a previously unseen Out-of-Distribution bone marrow dataset.}
    \label{figure:ood_detection_results_bone_marrow_dataset}
\end{figure}

Here, the performance of the conformal prediction methods in detecting a previously unseen bone marrow cytomorphology dataset is reported. \Cref{figure:ood_detection_results_bone_marrow_dataset} summarizes the performance. While the methods result in a larger set width when applied to the EfficientNet-B0 model, they do not show a similar trend in the rest of the models. So, similar to \Cref{section:results_ood_detction_noised_data}, we observe a model-dependent reaction to OOD data.

\section{Discussion}
\label{section:discussion}

Overall, we find that (1) LAC produces prediction sets with a lower mean width, resulting in higher mean precisions, while (2) RAPS achieves the highest mean recall on average, indicating that it generates the most prediction sets that align with expert annotations. These trends remain consistent across multiple values of $\alpha$. However, the performance of all the conformal prediction approaches was notably lower when compared to that of expert annotations on an exact match basis (see \Cref{table:set-by_set_comparison,figure:set-by_set_comparison}). This finding underscores a key limitation of conformal prediction methods. While in agreement with the existing literature \cite{angelopoulosuncertainty}, they consistently provide high coverage of the true class (see \Cref{table:cp_classification_performance_alpha_005}), they struggle to replicate expert-derived annotation sets accurately. Consequently, medical practitioners relying on conformal prediction must interpret the resulting prediction sets with caution. Although these methods ensure high coverage, they often introduce unlikely classes alongside the true class, which can contribute to misinformation. Users may be misled by the presence of the correct class within the prediction set, assuming that all predicted labels are meaningful when, in reality, many should be disregarded, as our results show.

Interestingly, most conformal prediction methods performed well in capturing aleatoric uncertainty, as evidenced by the alignment between their prediction set sizes and data ambiguity. However, their ability to capture epistemic uncertainty was poor. This was evident when we analyzed the changes in the prediction set widths as noise was incrementally added to the test dataset (see \Cref{figure:ood_detection_results}). Notably, only the ResNet50 model exhibited a corresponding increase in set width, indicating a model-dependent sensitivity to epistemic uncertainty. A similar model-dependent trend emerged when we evaluated the conformal prediction methods on a previously unseen dataset (see \Cref{figure:ood_detection_results_bone_marrow_dataset}), further highlighting the varying degrees of robustness across models in capturing epistemic uncertainty.

While the Fleiss Kappa inter-rater reliability analysis showed a fair agreement among the annotators, the test set size and the number of participant annotators limit our analysis of the conformal prediction methods. Our work, however, opens up an important avenue in validating conformal prediction for high-stakes areas such as medical image classification. For future work, we plan to extend the number of annotators and the number of test tiles.


\section{Conclusion}
\label{section:conclusion}

This work represents the first study to validate conformal prediction using annotation sets collected from multiple experts per input. While conventional conformal prediction evaluation metrics effectively assess the coverage of the true class within prediction sets, we extend these evaluations to measure how accurately these methods replicate expert-derived annotation sets. We evaluated three conformal prediction approaches applied to three deep learning models. Although these methods reliably cover the true class, they often simultaneously introduce unlikely classes within their prediction sets. This highlights a critical gap between coverage guarantees and practical usability in expert-driven domains. Our findings emphasize the need for cautious interpretation of conformal prediction outputs, particularly in high-stakes applications such as clinical decision-making. The key takeaway is that while conformal prediction can enhance uncertainty quantification, its outputs must be critically assessed to avoid potential misinformation.

\section*{Ethics statement}

Clinical trials have been conducted following the Helsinki Declaration and ICH Good Clinical Practice Guidelines. Ethical approval has been granted by the Technical University of Mombasa (TUM ERC EXT/001/2020). An agreement has been approved between the Kinondo Kwetu Trust Fund and the University of Helsinki, Karolinska Institute and Uppsala University (REF.OBN/C/22/04) as of 09/11/2022. The research for the clinical study in 2024 has been approved by the county government of Kwale, Kenya (REF: (CG/KWL/CECM/39VOL.1/ (55) as of 02/02/2024. 

\section*{Acknowledgments}

This work was partially supported by the Wallenberg AI, Autonomous Systems and Software Program (WASP) funded by the Knut and Alice Wallenberg Foundation, The Swedish e-Science Research Center, The Erling-Persson Foundation, The Swedish Research Council, Finska Läkaresällskapet, Medicinska Understödsföreningen Liv och Hälsa rf. and Wilhelm och Else Stockmanns stiftelse.


\section*{Author contributions}

\noindent \textit{Methodology, analysis, visualisation}: Misgina Tsighe Hagos\\
\textit{Manuscript draft}: Misgina Tsighe Hagos, Antti Suutala\\
\textit{Manuscript review and editing}:  Claes Lundström, Antti Suutala, Johan Lundin, Joar von Bahr, Milda Poceviciute\\
\textit{Model training}: Dmitrii Bychkov\\
\textit{Data preparation}: Antti Suutala\\
\textit{Data annotation platform}: Hakan Kücükel\\
\textit{Funding acquisition}: Nina Linder, Johan Lundin, Claes Lundström, Milda Poceviciute\\
\textit{Supervision}:  Nina Linder, Johan Lundin, Claes Lundström

\section*{Declaration of Interest Statement}

Johan Lundin is a co-founder and co-owner of Aiforia Technologies Plc. Claes Lundström is an employee of Sectra AB. No other disclosures were reported.

\bibliographystyle{elsarticle-num}

\IfFileExists{sample.bbl}{

}{
  \bibliography{sample.bib}  
}

\end{document}